\documentclass[structabstract,printer]{aa}

\usepackage{txfonts}
\usepackage{natbib}
\usepackage{graphicx}

\begin{document}

\title{Decaying dark matter in the Draco dwarf galaxy} 
\titlerunning{Decaying dark matter in Draco}


\author{Signe Riemer-S\o rensen\inst{1}\thanks{\email{signe@dark-cosmology.dk}}
  \and Steen H. Hansen\inst{1}}
\authorrunning{Riemer-S\o rensen and Hansen}

\institute{Dark Cosmology Centre, Niels Bohr Institute, University of Copenhagen, Juliane Maries Vej 30, DK-2100 Copenhagen, Denmark \thanks{The Dark Cosmology Centre is funded by the Danish National Research Foundation}}

\date{Received 2 November 1992 / Accepted 7 January 1993}

\abstract{The sterile neutrino is an excellent dark matter candidate, which can be searched for in a wide range of astrophysical sites. It has previously been shown that the optimal search strategy is to consider dwarf galaxies belonging to the Milky Way.}
{We search for line emission from decaying dark matter.} {We analyse publicly available Chandra X-ray observations of the dwarf galaxy Draco.}
{The Draco and blank sky (background) spectra are nearly identical in shape, which allows us to conclude: i) Dwarf spheroidals are ideal for studying dark matter X-ray emission since the baryonic noise is impressively low, ii) there is very little room for line emission, which leads to constraints in the mass-mixing angle parameter space of the sterile neutrino. We compare the standard flux derivation method to a very conservative rebinning approach. The resulting constraints are strongly dependent on the chosen method.}
 {}

\maketitle 
\section{Introduction}
The sterile neutrino is a strong particle candidate for dark matter (DM). With just three sterile neutrinos (gauge singlets), one easily obtains the correct abundance of dark matter, a very simple explanation for the observed flavor oscillations of the active 
neutrinos as well as their mass splitting, and, in addition, is a natural explanation for the baryon asymmetry \citep{Dodelson:93, Shi:98, Dolgov:00, Abazajian:01a, Asaka:05a, Asaka:05b}. The underlying particle model, called the $\nu$MSM, is described in detail in a number of papers \citep{Asaka:05a, Asaka:05b, Asaka:06c, Shaposhnikov:06b, Gorbunov:07a, Laine:08a, Shaposhnikov:08a}, and also in the excellent review by \citet{review}. Additionally, the sterile neutrino may have interesting effects on a range of different astrophysical objects, including as an explanation for pulsar kick velocities,  facilitating core collapse supernova explosions, affecting early star formation, reionization and structure formation, or assisting inflation \citep{Kusenko:97,Fryer:05,Hidaka:06, Biermann:06,Hansen:04,Mapelli:06,Bezrukov:07,Shaposhnikov:06c,Kusenko:07,Petraki:07,Petraki:07b,Boyanovsky:08,Gorbunov:08} (see \citet{review} for an extensive list of references).

The large range of astrophysical objects affected by the sterile neutrino allows for independent measurements of these new particles.
The two free parameters for the lightest sterile neutrino, the DM particle,  is its mass, $m_s$, and mixing angle with the active neutrinos, ${\rm sin}^2(2\theta)$, and various observations have already excluded large parts of this parameter space.

Structure formation will be changed if the DM particle is warm as compared to cold, and for instance Lyman-$\alpha$ absorption in quasar spectra allows one to place lower bounds on the DM particle mass \citep{Hansen:02, Viel:06, Viel:05, Seljak:06}. The most recent Lyman-$\alpha$ analysis \citep{Boyarsky:08c,Boyarsky:08e} finds that for all masses above $m_s=2$~keV, there exist parts of the parameter space in agreement with all observations. The originally proposed particle production method, non-resonant production \citep{Dodelson:93}, is already ruled out observationally, and only the resonant production (RP) \citep{Shi:98}, with a large initial lepton asymmetry, is in agreement with all observations. This lepton asymmetry can be constrained directly from nucleosynthesis \citep{Serpico:05, Dolgov:02}. Given this limit on the lepton asymmetry the RP needs a sufficiently large mixing angle to make the abundance of the sterile neutrino as large as the observed DM abundance. This gives a strong lower limit on the mixing angle, which is approximately ${\rm sin}^2(2\theta) \ge 10^{-12}$ for $m_s=2$~keV\citep{Laine:08a}.

A different, but very firm, lower boundary on the mass is obtained through the phase space density of nearby dwarf galaxies. The Tremaine-Gunn bound \citep{Tremaine:79} gives a model independent boundary of roughly $0.4$~keV \citep{Boyarsky:08a}. This boundry can be strengthened if the production method is known, and for the RP the boundary is approximately 1~keV \citep{Boyarsky:08a}.

The first constraints on the mixing angle came from the non-observation of decay lines \citep{Dolgov:00}. This is because the sterile neutrino decays through a 1-loop into a photon and an active neutrino, with a decay rate of \citep{Pal:81,Barger:95}:
\begin{equation}
\Gamma  = 1.4 \times 10^{-22}  \, {\rm sin}^2 (2\theta) \,\left( \frac{m_s}{{\rm keV}} \right) ^5 \, {\rm sec}^{-1} \, .
\end{equation}
Since the DM decay almost at rest, the decay line is very narrow and easily searched for in X-ray and soft gamma ray observations \citep{Dolgov:00,Riemer:07,Abazajian:06b,Riemer:06,Boyarsky:06c, Boyarsky:06d, Abazajian:01b, Boyarsky:06e, Boyarsky:07,Boyarsky:07b,Yuksel:07,Watson:06,Loewenstein:08}.

The main problem with X-ray observations is the baryonic background, and one therefore needs to examine sites with very few baryons. Such systems already considered include the dark blobs in merging clusters \citep{Hansen:02,Riemer:07,Abazajian:06b} or simply looking out through our own Milky Way halo \citep{Riemer:06,Boyarsky:06c}. However, it was demonstrated that the optimal place to search for the decay line of the sterile neutrino is from the dwarf galaxies of our Milky Way such as Ursa Minor, Draco etc. \citep{Boyarsky:06c,Boyarsky:06d}. These dwarfs are nearby, they are X-ray quiet, have a high central DM density, and at the same time we have fairly accurate mass models from optical observations \citep{Gilmore:2007, Wilkinson:2004, Strigari:08, Strigari:07}.

In this paper, we analyse {\it Chandra} X-ray data of Draco. The Draco and blank sky spectra are almost identical and leave no significant room for line emission from decaying DM. The absence of any line signal is used to derive constraints in the mass-mixing angle parameter space of the sterile neutrino.

\section{The observations} \label{sec:technical}
\subsection{X-ray analysis}
There exist two public {\it Chandra} observations of Draco with observation ids 9568 (24.8~ks) and 9776 (12.2~ks) in the NASA HEASARC archive\footnote{http://heasarc.gsfc.nasa.gov/docs/archive.html}. Before extracting spectra, we reprocessed the data with the newest versions of CIAO (4.1) and CALDB (4.1) following standard procedures \citep{Fruscione:2006}. A total of 12 evenly distributed point sources were removed and the light curve cleaned, reducing the total exposures to 20.3~ks (9568) and 12.0~ks (9776)). The central part of Draco is situated on the S3 chip. The spectra were extracted from a square region of $(7.6~\arcmin)^2 \approx (0.18{\rm kpc})^2$, avoiding the edges of the chip. The response matrices were generated and the spectra and response matrices of the two observations were combined using {\it ftools}\footnote{{\it ftools}, HeaSoft, http://heasarc.gsfc.nasa.gov/docs/\\software/ftools/ftools\_menu.html} (justified by the very similar count rates in the two otherwise identical observations). For background subtraction we extracted the spectrum of the identical region of the chip from the blank sky data provided with CIAO. 

\begin{figure}[tbp]
\centering
	\includegraphics[angle=270,width=8.5cm]{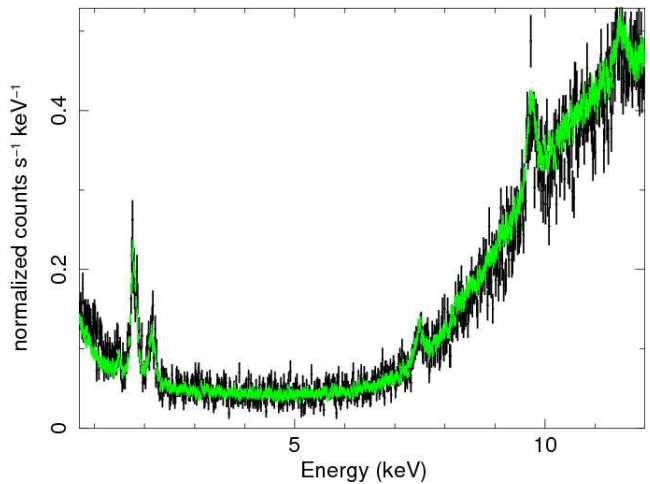}
	\caption{The spectrum of the central $(7.6~\arcmin)^2$ (black) and the corresponding normalized blank sky spectrum (green).}
	\label{fig:spectrum}
\end{figure}

The Draco and blank sky spectra are shown in Figure \ref{fig:spectrum}, where it is clearly seen that they are almost identical, with very little room for a dark matter signal. 

The normalization of the blank sky spectrum was determined from the average count rate in the 4-6~keV interval for both blank sky and observation. This interval was chosen because both spectra are very flat and well calibrated here. Additionally, no thermal gas emission is expected at these energies, as Draco is a small and cold structure. The uncertainty of the average count rate is less than 0.5\%. Normalising at higher energies gives slightly different background levels e.g. 10-12~keV gives a 3.5\% lower level, and 12-14~keV (which is usually used for galaxy cluster observations) gives a 0.5\% higher level. However, both intervals have higher uncertainties on the normalisation due to the large scatter. 

\begin{figure}[tbp]
\centering
	\includegraphics[width=8.5cm]{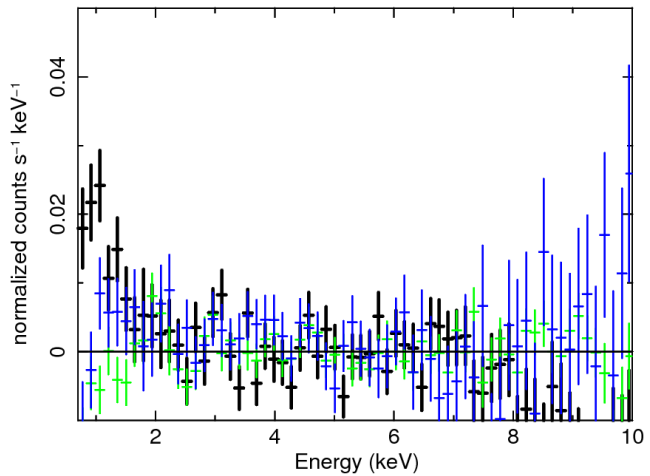}
	\caption{The background subtracted spectra of the S3 (center of Draco, fat black), S2 ($8.6~\arcmin$ from center, green), and S3 ($17.3~\arcmin$ from center blue) chips rebinned for visualisation. In the case of any observable signal from dark matter, there will be a radial dependence, like the one around 1~keV, where the signal disappears far from the center.}
	\label{fig:3spec}
\end{figure}

The background-subtracted Draco spectrum is shown in Figure \ref{fig:3spec} (black). Apart from a small signal $\approx$1~keV, the spectrum is effectively zero. From this we see, i) that dwarf spheroidals have no significant baryonic X-ray emission, ii) there is virtually no room for a dark matter decay line.

After the background subtraction, the only possible signal is at low energies ($\approx 0.7-1.5$~keV). The best fitting Gaussian has a width of $\ge0.5$~keV, which is an order of magnitude larger than the width of the instrumental resolution at 1~keV. To check for any radial dependence in the signal, we extracted spectra from similar regions of the S2 and S1 chips centered $\approx 8.6~\arcmin$ and $\approx17.3~\arcmin$ from the center of Draco. The three spectra are compared in Figure \ref{fig:3spec}. The low energy ``bump'' around 1~keV is not clearly present in the outer radial bins. It is tempting to believe that this is a signal proportional to the mass within the field of view. Nonetheless, the excess is not very significant and is too wide to be a monochromatic emission line. Most astrophysical backgrounds such as stellar and thermal radiation will also follow the mass distribution of Draco.

The upper limit on the flux was derived in the 0.7-10~keV interval using the spectral fitting package Xspec \citep[version 12.4,][]{xspec} following different approaches. The flux derivations are based on the dark matter in Draco being practically at rest \citep[$v/c \approx 10^{-4}$,][]{Gilmore:2007} so the line broadening due to internal motion is negligible compared to the instrumental resolution. The instrumental resolution is given by: $E_{FWHM}=0.012E_\gamma+0.12$keV \citep{POG}. 

\begin{figure}[tbp]
\centering
	\includegraphics[angle=270,width=8.5cm]{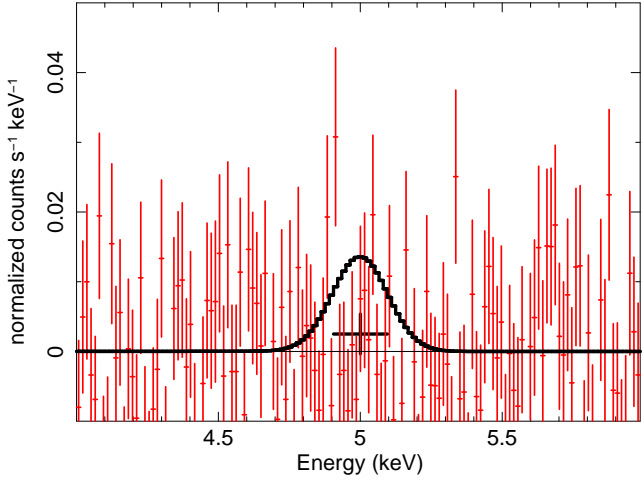}
	\caption{The rebinning approach for $E_\gamma=5$~keV. All data within the interval $[E_\gamma-E_{FWHM}/2:E_\gamma+E_{FWHM}/2]$ was rebinned to one bin (central data point). The width of the Gaussian is given by the resolution and the maximum by the $3\sigma$ upper limit of the rebinned data.}
	\label{fig:binning}
\end{figure}

One employed approach is to rebin the data with a binwidth corresponding to the instrumental resolution. At a given energy, $E_\gamma$, all bins within the interval $[E_\gamma-E_{FWHM}/2:E_\gamma+E_{FWHM}/2]$ were rebinned to give a single bin value and uncertainty (fat data point in Figure \ref{fig:binning}). The possible line emission was defined as a Gaussian with the width given by the resolution ($E_{FWHM}=2.35E_{\sigma}$ for a Gaussian distribution) and maximum at the $3\sigma$ upper limit of the rebinned data. The upper limit on the flux was determined from the Gaussian over the same interval as the data were rebinned.

\begin{figure}[tbp]
\centering
	\includegraphics[angle=270,width=8.4cm]{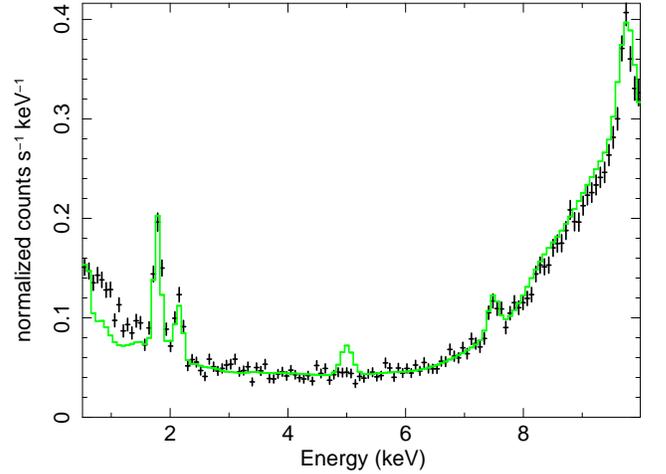}
	\caption{The Draco spectrum and the background model (green line) including the potential line at $E_\gamma=5$~keV. For visualisation, the line normalisation has been increased by a factor of 10.}
	\label{fig:gauss}
\end{figure}

Another approach was to model the background instead of subtracting it \citep[same method as discussed in][]{Boyarsky:06c}. The background was fitted with a continuum consisting of an exponential and 10 Gaussians. The fit gave a reduced $\chi^2$ of $1.3$ (for $637$ d.o.f.). The line emission was added to the model as a Gaussian with the width given by the instrumental resolution (see Figure \ref{fig:gauss}). The total model was fitted to the data with the normalisation of the Gaussian as the only free parameter. The flux was determined exclusively from the Gaussian with the normalisation set to the $3\sigma$ upper limit from the fit.

\subsection{Observed mass}
The dark matter mass within the field of view is determined from the density profile derived from photometric and kinematic data by \citet{Gilmore:2007}. The density profile in the $0.1-0.5$~kpc range is well described by $\rho(r)=\rho_0 (1+(r/r_0)^{1.5})^{-2.25}$ where $\rho_0 = 0.65\times 10^9 M_\odot$~kpc$^{-3}$ and $r_0 = 0.28$~kpc. We extrapolated this profile to the observed range of $0-0.18$~kpc and integrated along the line of sight to determine the mass within the field of view. The total dark matter mass of Draco within $0.5$~kpc is $\ge 6\times10^{7}~M_\odot$ \citep{Gilmore:2007, Strigari:08} of which $\approx 6.7\times10^{6}~M_\odot$ is within the observed square. The statistical uncertainty on the mass is very small, but the systematics are very hard to determine \citep{Gilmore:2007,Boyarsky:06d}. 

\section{Results}
\begin{figure}[tbp]
\centering
	\includegraphics[width=8.5cm]{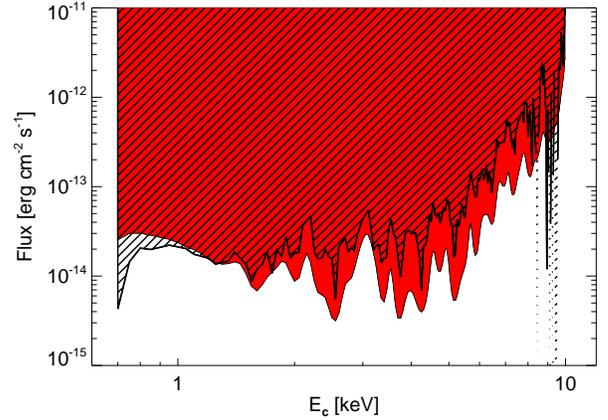}
	\caption{The obtained upper limits on the line emission flux. The modelling Gaussian approach (solid red) gives stronger constraints than the conservative rebinning approach (black hatches).}
	\label{fig:flux}
\end{figure}

Figure \ref{fig:flux} shows the obtained upper limits on line emission flux as a function of photon energy (coloured areas are excluded). These constraints are very general and apply to all dark matter candidates with a monochromatic line emission. The vertical lines in the rebinned approach constraints are single energies, where the upper limit on the flux becomes exactly zero for a $3\sigma$ upper limit on the flux (using this approach with the chosen background normalisation). If we require e.g. $5\sigma$, they will disappear.

\begin{figure}[tbp]
\centering
	\includegraphics[width=8.5cm]{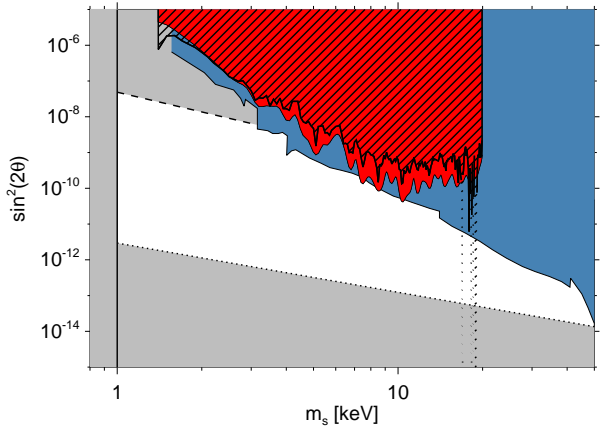}
	\caption{The obtained constraints in the $m_s - \sin^2(2\theta)$ parameter space for sterile neutrinos of the Majorana type. The modelling Gaussian approach (solid red) gives stronger constraints than the conservative rebinning approach (black hatched).  Earlier constraints \citep[taken from][and references therein]{review}: Phase space considerations (grey below 1~keV), nucleosynthesis (grey region below dotted line), non-resonant production giving $\Omega>\Omega_{DM}$ (grey region above dashed line), earlier observational constraints (blue).}
	\label{fig:msin}
\end{figure}

The fluxes are converted to constraints in the $m_s - \sin^2(2\theta)$ parameter space for sterile neutrinos of the Majorana type, assuming the sterile neutrinos to account for all of the dark matter in Draco \citep{Riemer:06,Boyarsky:06c}:
\begin{eqnarray}
\sin^2(2\theta) &\leq& 1 \times 10^{18} \left(\frac{F_{det}}{{\rm erg cm^{-2} s^{-1}}}\right) \times \left(\frac{m_s}{{\rm keV}}\right)^{-5} \left[\frac{(M_{fov}/M_\odot)}{(D_L/{\rm Mpc})^2}  \right]^{-1} \, .
\end{eqnarray}
The distance to Draco is $80$~kpc \citep{Gilmore:2007}. The resulting conservative constraints are shown in Figure \ref{fig:msin} (red and black hatched) compared to earlier constraints from galaxy clusters, galaxies, and the Milky Way halo \citep[blue, ][and references therein]{review}. The overlap with earlier constraints is an excellent confirmation of the excluded parameter space, since the Draco constraints are complementary and independent of earlier constraints. Even with the relatively short exposure, the constraints are good due to the almost perfect similarity of the spectra of Draco and the background. 
The difference between the Gaussian modelling (red) and the rebinning (black hatched) approaches is clearly visible, with the Gaussian modelling method providing stronger constraints at most energies than the more conservative rebinning method. 

\section{Future observations?}
The limiting factor of the constraints are the uncertainties in the observations and blank sky, which can be improved by increasing the exposure time. The statistics improve with the square root of the exposure time so a 100~ks observation improves the constraints by a factor of $\approx 2$. In order to exclude the entire currently allowed region, the needed exposure time for {\it Chandra} or similar instruments becomes unreasonably high and a different instrumental approach is needed \citep{herder}. One option could be gratings, but unfortunately gratings are not useful for extended sources because of smearing due to the ambiguity between energy and incident angle of the photon \citep{Riemer:07}. 

\section{Summary}
We have searched for line emission from decaying dark matter particles in {\it Chandra} X-ray observations of the Milky Way dwarf galaxy Draco. The Draco and blank sky background spectra have a nearly identical shape. This confirms dwarfs as ideal for studying dark matter emission since the baryonic contamination is close to zero. No obvious line signal is found, which leads to conservative constraints in the $m_s - \sin^2(2\theta)$ parameter space for sterile neutrinos. Longer exposures are needed to reduce the statistical uncertainty and improve the constraints.

\begin{acknowledgements}
We would like to thank Darach Watson for inspiring discussions and comments, and Alexander Kusenko, Michael Loewenstein, Alexey Boyarsky, Oleg Ruchayskiy, and Kristian Pedersen for useful suggestions. 
\end{acknowledgements}

\bibliographystyle{aa} 
\bibliography{./2430_refs.bib} 

\end{document}